%% file: eprint.tex
\documentclass[12pt]{article}
\usepackage{graphicx}

\newcommand\pubnumber{}
\newcommand\pubdate{\today}

\def\GHMC{Gunma University Heavy Ion Medical Center\\
3-39-22 Showa-machi, Maebashi, Gunma 371-8511, Japan}

\def\Title#1{\begin{center} {\Large #1 } \end{center}}
\def\Author#1{\begin{center}{ \sc #1} \end{center}}
\def\Address#1{\begin{center}{ \it #1} \end{center}}

\newcommand\pubblock{\rightline{\begin{tabular}{l} \pubnumber \\
      \pubdate  \end{tabular}}}
\newenvironment{Abstract}{\begin{quotation}  }{\end{quotation}}
\newenvironment{Presented}{\begin{quotation} \begin{center} 
             PRESENTED AT\end{center}\bigskip 
      \begin{center}\begin{large}}{\end{large}\end{center} \end{quotation}}
\def\Acknowledgements{\bigskip  \bigskip \begin{center} \begin{large}
             \bf ACKNOWLEDGEMENTS \end{large}\end{center}}

\input econfmacros.tex

\begin{document}
\begin{titlepage}
\pubblock

\vfill
\Title{Response of SOI image sensor to therapeutic carbon ion beam}
\vfill
\Author{ Akihiko Matsumura}
\Address{\GHMC}
\vfill
\begin{Abstract}
Carbon ion radiotherapy is known as a less invasive cancer treatment.
The radiation quality is an important parameter to evaluate
the biological effect and the clinical dose from the 
measured physical dose.
The performance of SOPHIAS detector, which is the SOI 
image sensor having a wide dynamic range and large 
active area, was tested by using therapeutic carbon ion beam
at Gunma University Heavy Ion Medical Center (GHMC).
It was shown that the primary carbon and secondary particles
can be distinguishable by SOPHIAS detector. 
On the other hand, a LET dependence was observed especially
at the high LET region.
This phenomenon will be studied by using the device simulator
together with Monte Carlo simulation.
\end{Abstract}
\vfill
\begin{Presented}
International Workshop on SOI Pixel Detector (SOIPIX2015)\\
Tohoku University, Sendai, Japan, 3-6, June, 2015.
\end{Presented}
\vfill
\end{titlepage}
\def\thefootnote{\fnsymbol{footnote}}
\setcounter{footnote}{0}

\section{Introduction}
Carbon ion radiotherapy has attracted increasing attention recently 
as a minimally invasive cancer treatment due to its superior 
characteristics, such as a high linear energy transfer (LET) and a
better dose localization~\cite{Tsujii,Ohno}. 
The biological dose can be obtained for the product of the physical
dose and relative biological effectiveness (RBE) which depends on
the LET~\cite{Kanai}. 
The LET distribution is therefore a significant information
to evaluate the clinical dose by using the measured physical dose.

\begin{figure}[htb]
\centering
\begin{tabular}{cc}
\begin{minipage}[t]{0.5\hsize}
\includegraphics[width=7.5cm]{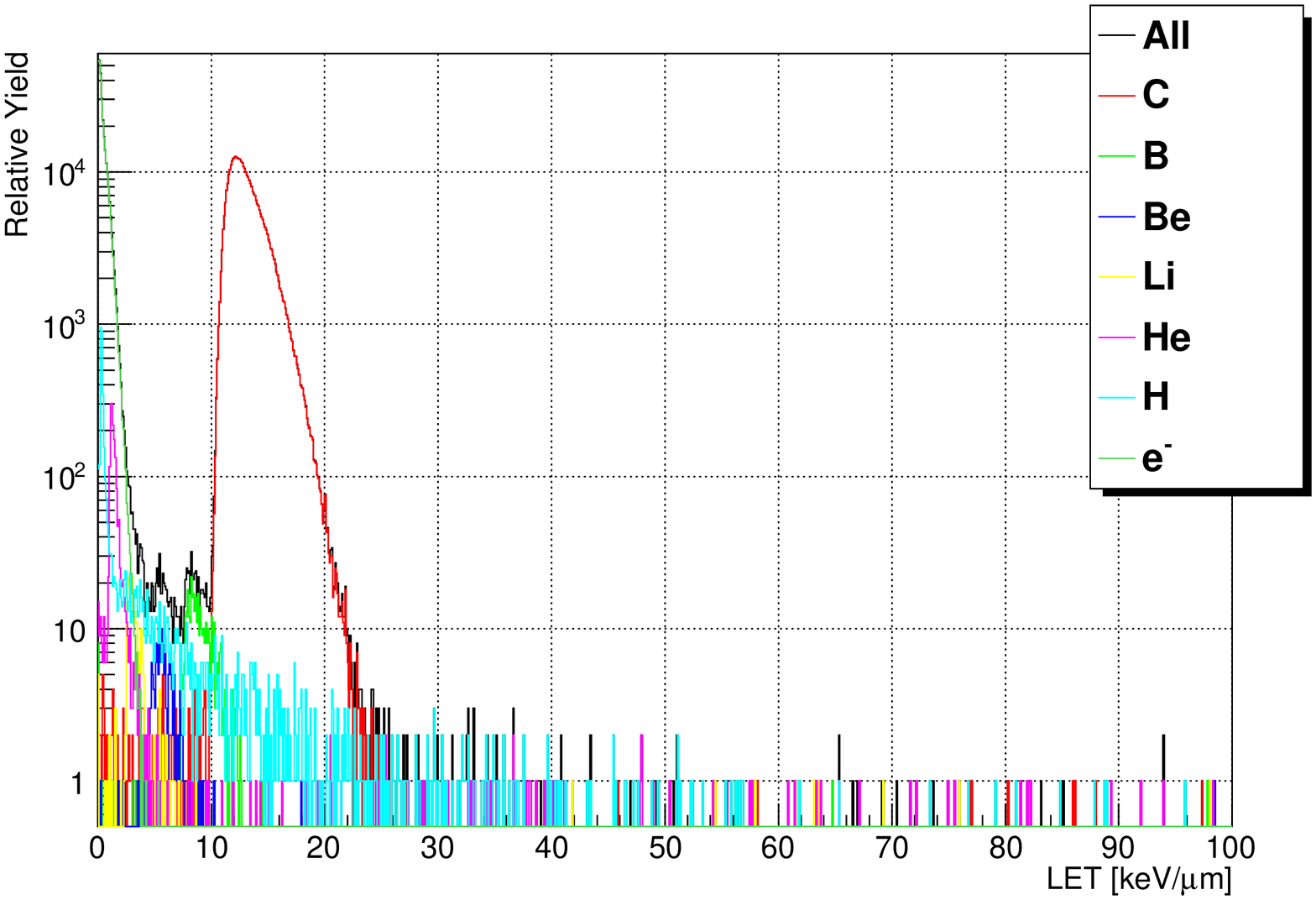}
\end{minipage}
\begin{minipage}[t]{0.5\hsize}
\includegraphics[width=7.5cm]{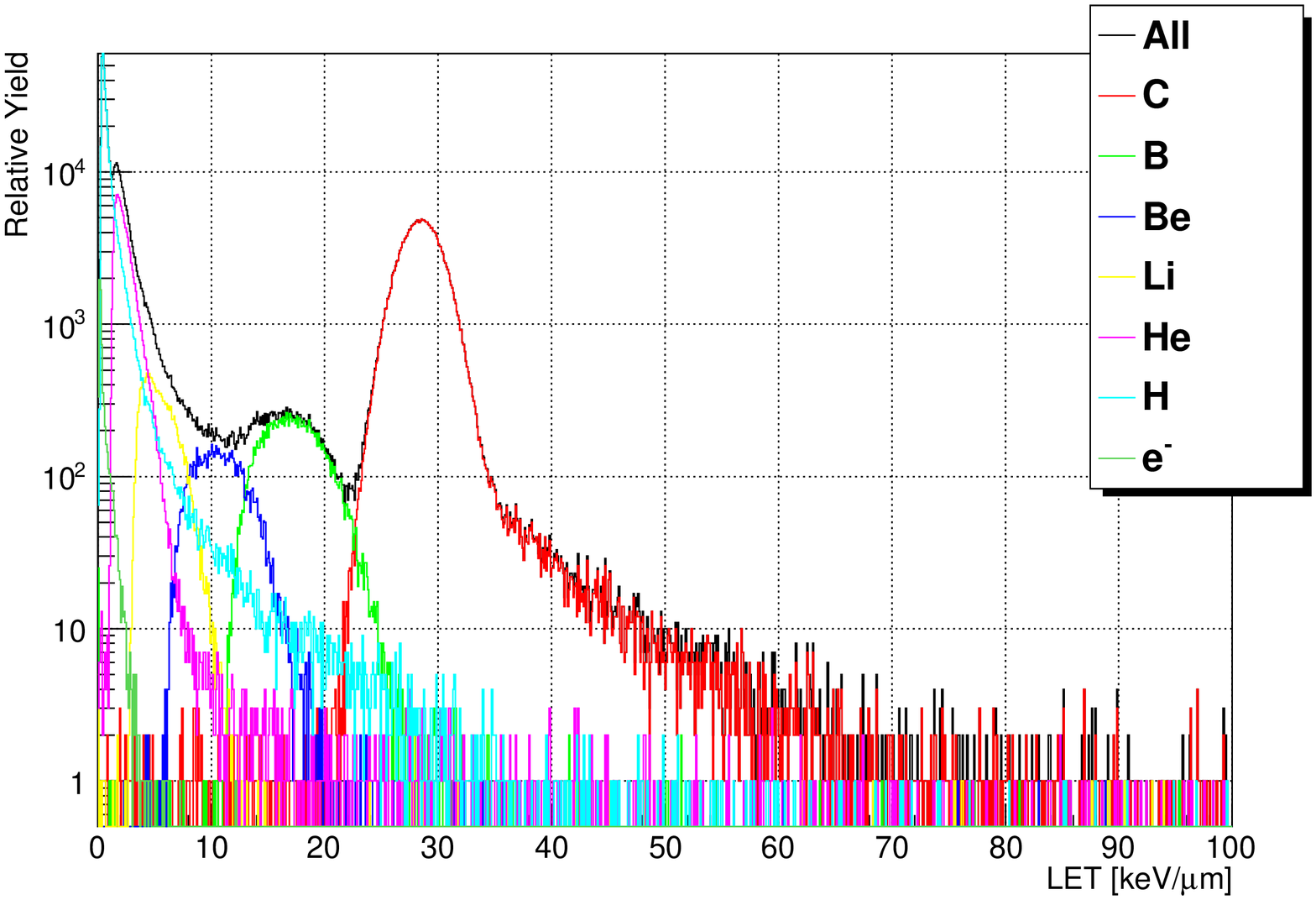}
\end{minipage}
\end{tabular}
\caption{Calculated LET distribution in the water target by 
Monte Carlo simulation (GEANT4). Left and right represent the LET
distribution at depth = 0 mm and depth = 120 mm (in front of 
Bragg peak), respectively.}
\label{fig:GEANT4}
\end{figure}

A conventional broad beam method is applied to cancer treatment
with the therapeutic carbon beam at Gunma University Heavy Ion Medical
Center (GHMC)~\cite{Torikoshi,Ohno2}. 
Accelerated carbon ions are passively scattered by
various beam line devices to form three-dimensionally dose distribution.
Many secondary radiations are generated when primary carbon ions
pass through various beam line devices. Thus the radiation quality
of the patient-specific irradiation field is very complicated.
Figure \ref{fig:GEANT4} shows the calculated LET distribution 
in the water by Monte Carlo simulation, GEANT4. 
In this simulation, primary carbon ions 
of 290 MeV/u pass through scatterer made of lead with a thickness 
of 2.3 mm and then enter water target. This result shows the amount of 
secondary radiations increase while that of primary carbons decrease in
the water. A detailed LET measurement of therapeutic carbon ion beam
offers new knowledge and information about carbon ion radiotherapy.

SOI image sensor has a high spatial resolution and is relatively 
resistance to single event effects~\cite{Arai}. Especially, the
SOPHIAS detector, which has a wide dynamic range and a large active
area~\cite{Hatsui}, could be a breakthrough dosimetric device in
carbon ion radiotherapy. In this report, the experimental result
with regard to the response of SOI image sensor to therapeutic
carbon ion beam will be presented.

\section{Experiment}

\begin{figure}[htb]
\centering
\includegraphics[width=7cm]{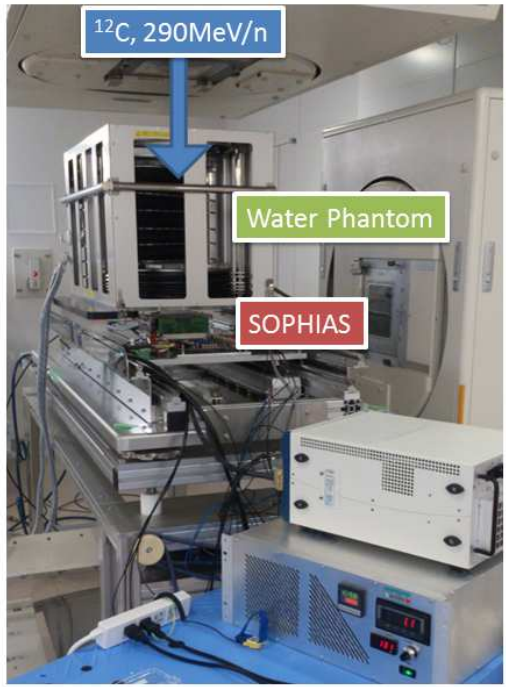}
\caption{Experimental setup.}
\label{fig:Setup}
\end{figure}

The beam test of SOPHIAS detector was performed at vertical port
of treatment room B in GHMC. Figure \ref{fig:Setup} shows the
experimental setup. 
A water phantom was located at the upstream position 
of SOPHIAS detector. 
The thickness of water can be changed by remote
control to measure the LET distribution at any depth easily. 
Mono-energetic carbon beam of 290 MeV/u with a nominal field size of
110 mm was used and the multi-leaf collimator was adjusted to protect
the circuit. 
Beam intensity was reduced to 1/40 of treatment condition 
by using attenuator and adjusting the chopper parameters
to suppress the multiple events in a single pixel.
It was confirmed that the performance of the beam monitor which 
controls the irradiation dose kept stable even under the
low beam intensity.

The back bias voltage and exposure time of SOPHIAS detector were set to
200 V and 100 $\mu$s, respectively. The sensor was cooled to
around -5 $^\circ$C by Peltier device. The active area of SOPHIAS with
a pixel spacing of 30 $\mu$m is 26.73 mm $\times$ 64.71 mm
( 891 pixel $\times$ 2157 pixel ). 
In this study, the data for 2/3 of active area was taken 
due to the limitation of read-out system.
As a reference, the data under the same condition except for the
beam intensity was taken by the parallel plane ionization chamber 
(PTW 34045).

\begin{figure}[htb]
\centering
\includegraphics[width=11cm]{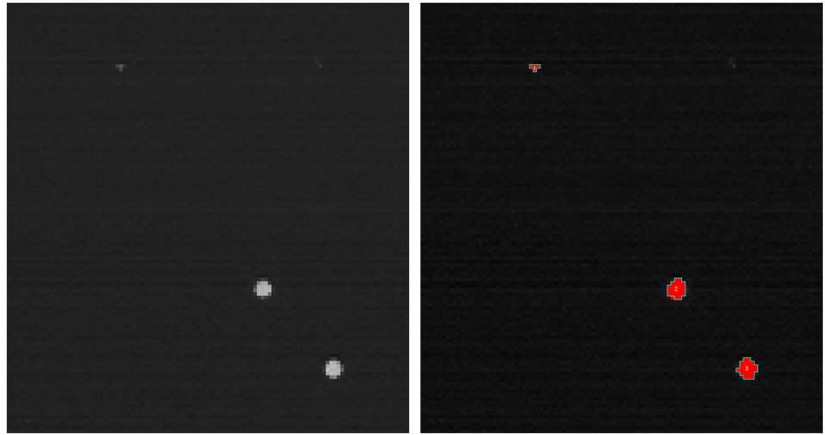}
\caption{Example of image analysis by ImageJ. Left shows the raw image
frame obtained by SOPHIAS detector. Right represents the result of
clustering shown in red.}
\label{fig:ImageJ}
\end{figure}

Obtained images were analyzed with the image processing software, 
Image J v1.47. The signal above a threshold of 0.02 V was clustered
as a "{\it Particle}". The "{\it Area}" and the "{\it Sum}" of pixel
values were then extracted for each {\it Particle}. 
Figure \ref{fig:ImageJ} shows the example of image analysis.

\section{Results and Discussion}

\begin{figure}[htb]
\centering
\begin{tabular}{cc}
\begin{minipage}[t]{0.5\hsize}
\includegraphics[width=7.5cm]{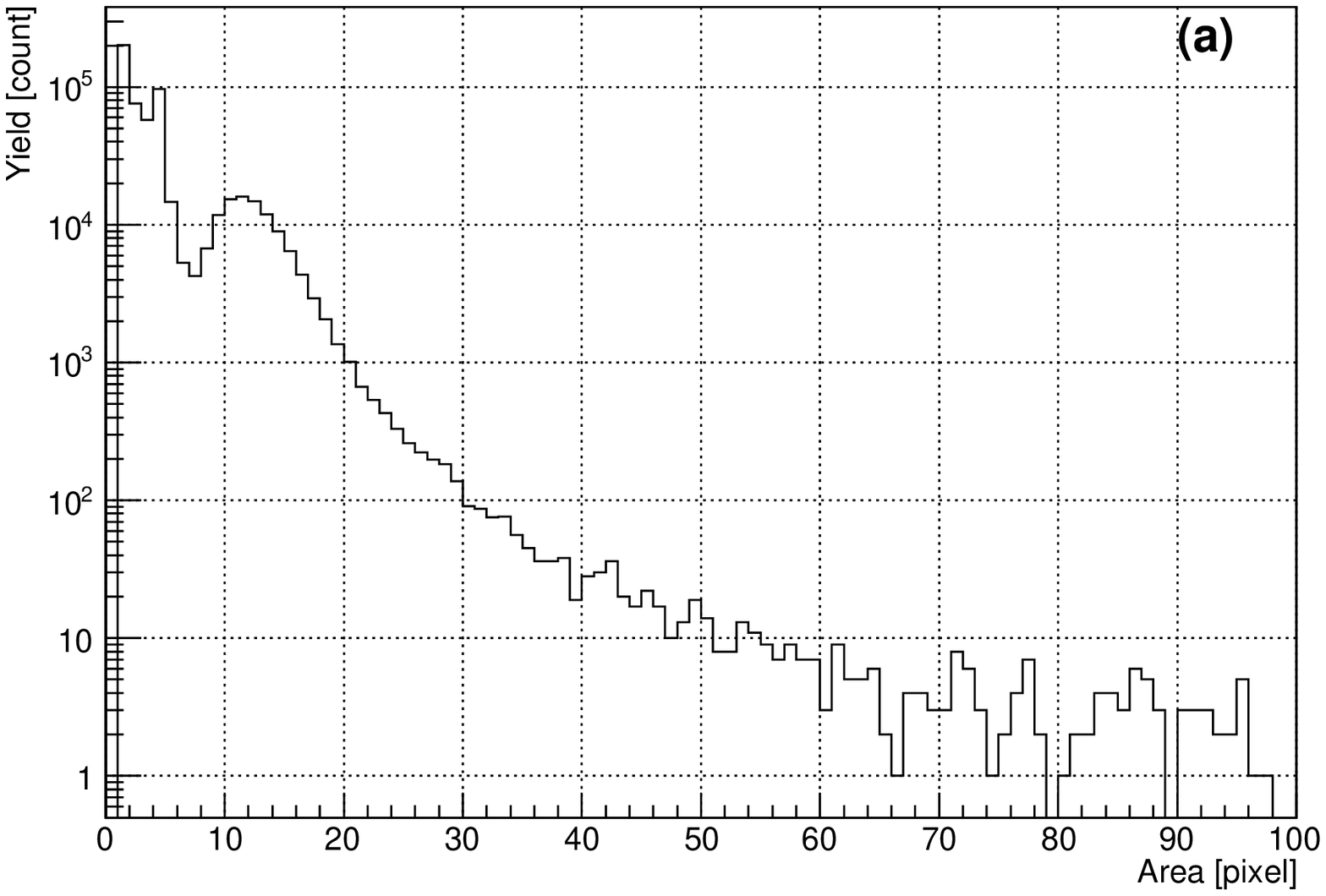}
\end{minipage}
\begin{minipage}[t]{0.5\hsize}
\includegraphics[width=7.5cm]{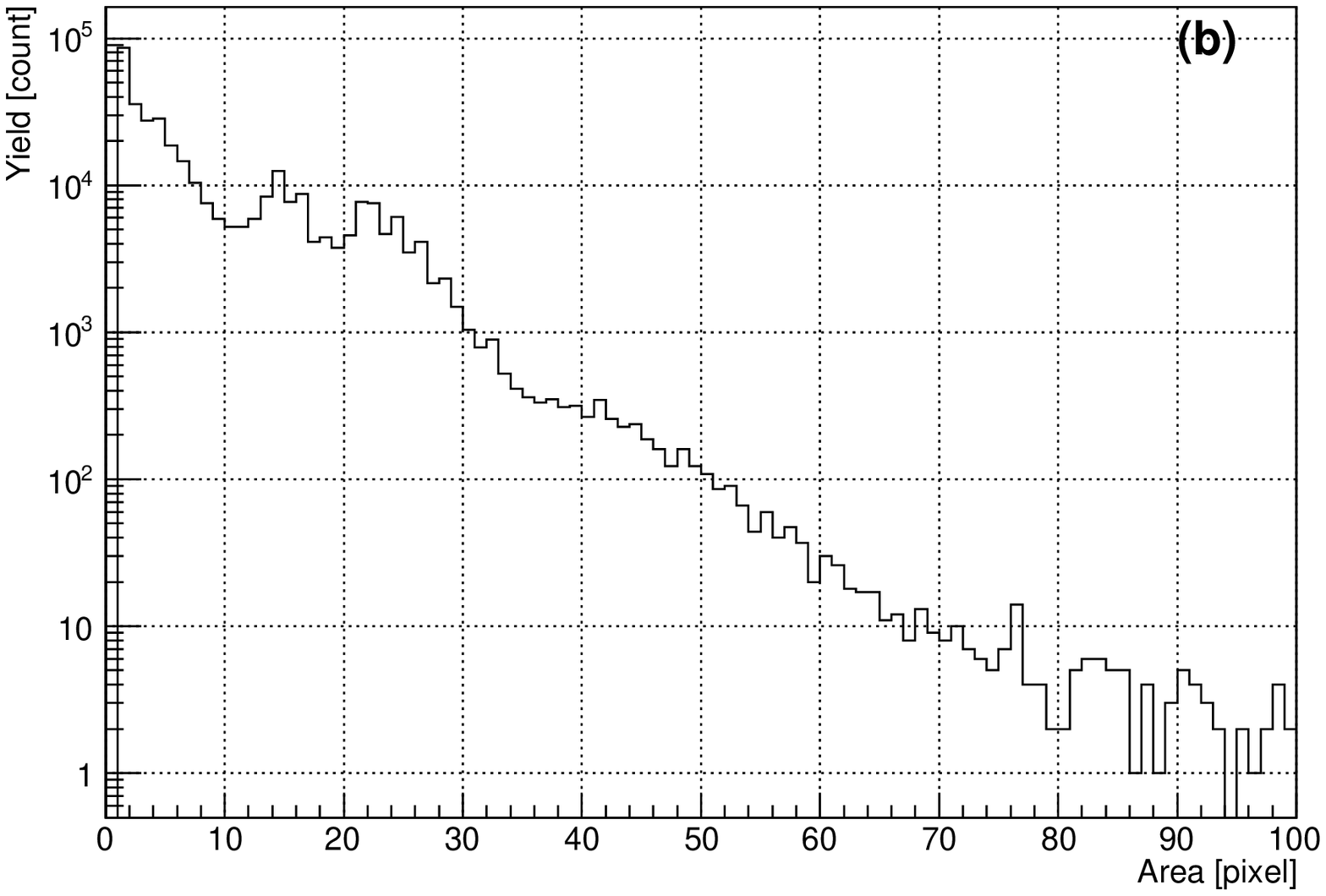}
\end{minipage}
\end{tabular}
\begin{tabular}{cc}
\begin{minipage}[t]{0.5\hsize}
\includegraphics[width=7.5cm]{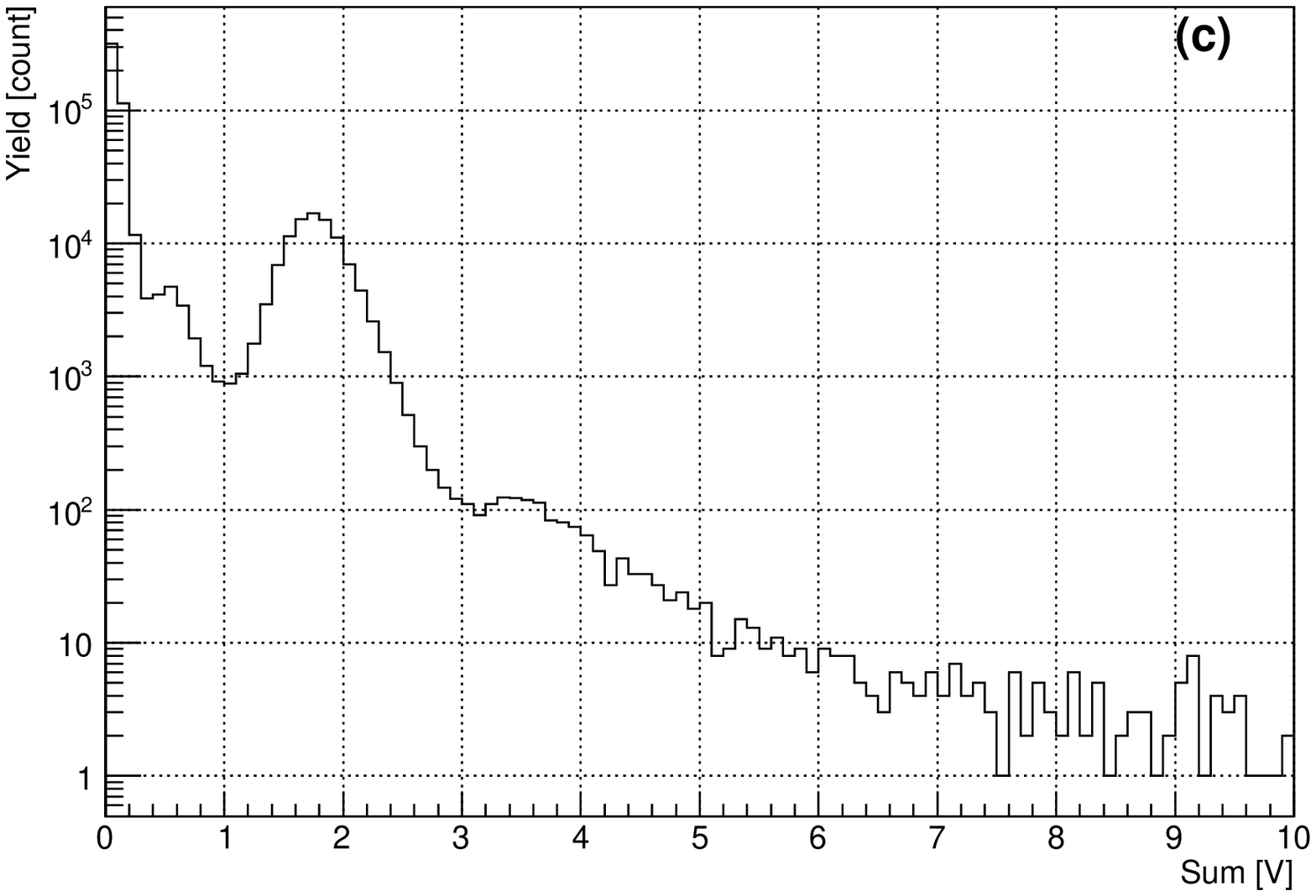}
\end{minipage}
\begin{minipage}[t]{0.5\hsize}
\includegraphics[width=7.5cm]{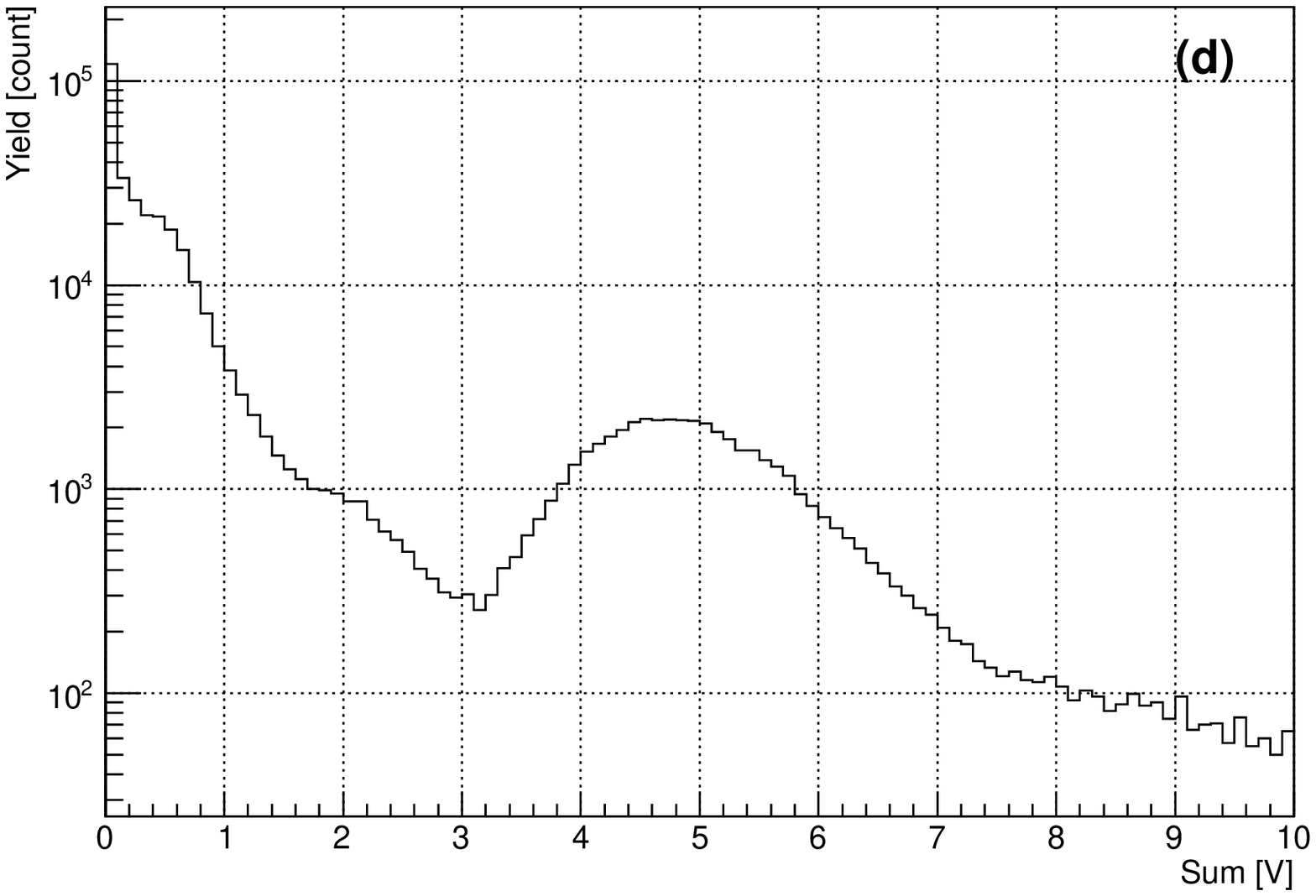}
\end{minipage}
\end{tabular}
\caption{Obtained {\it Area} and {\it Sum} distributions. 
(a) {\it Area} distribution at a depth of 0 mm (plateau region).
(b) {\it Area} distribution at a depth of 112 mm (Bragg peak).
(c) {\it Sum} distribution at a depth of 0 mm (plateau region).
(d) {\it Sum} distribution at a depth of 112 mm (Bragg peak).
}
\label{fig:1DHist}
\end{figure}

Figure \ref{fig:1DHist} represent the {\it Area} and {\it Sum}
distribution at plateau and Bragg peak region. 
No significant peak structure can be found 
in {\it Area} histograms though the {\it Area} value increase 
slightly at Bragg peak. 
On the other hand, a major peak can be observed in the {\it Sum}
histograms. The {\it Sum} value of the peak increase and the number of
events decrease at Bragg peak region. These tendencies seem to be
similar to the simulation result shown in Figure \ref{fig:GEANT4}.

\begin{figure}[htb]
\centering
\begin{tabular}{cc}
\begin{minipage}[t]{0.5\hsize}
\includegraphics[width=7.5cm]{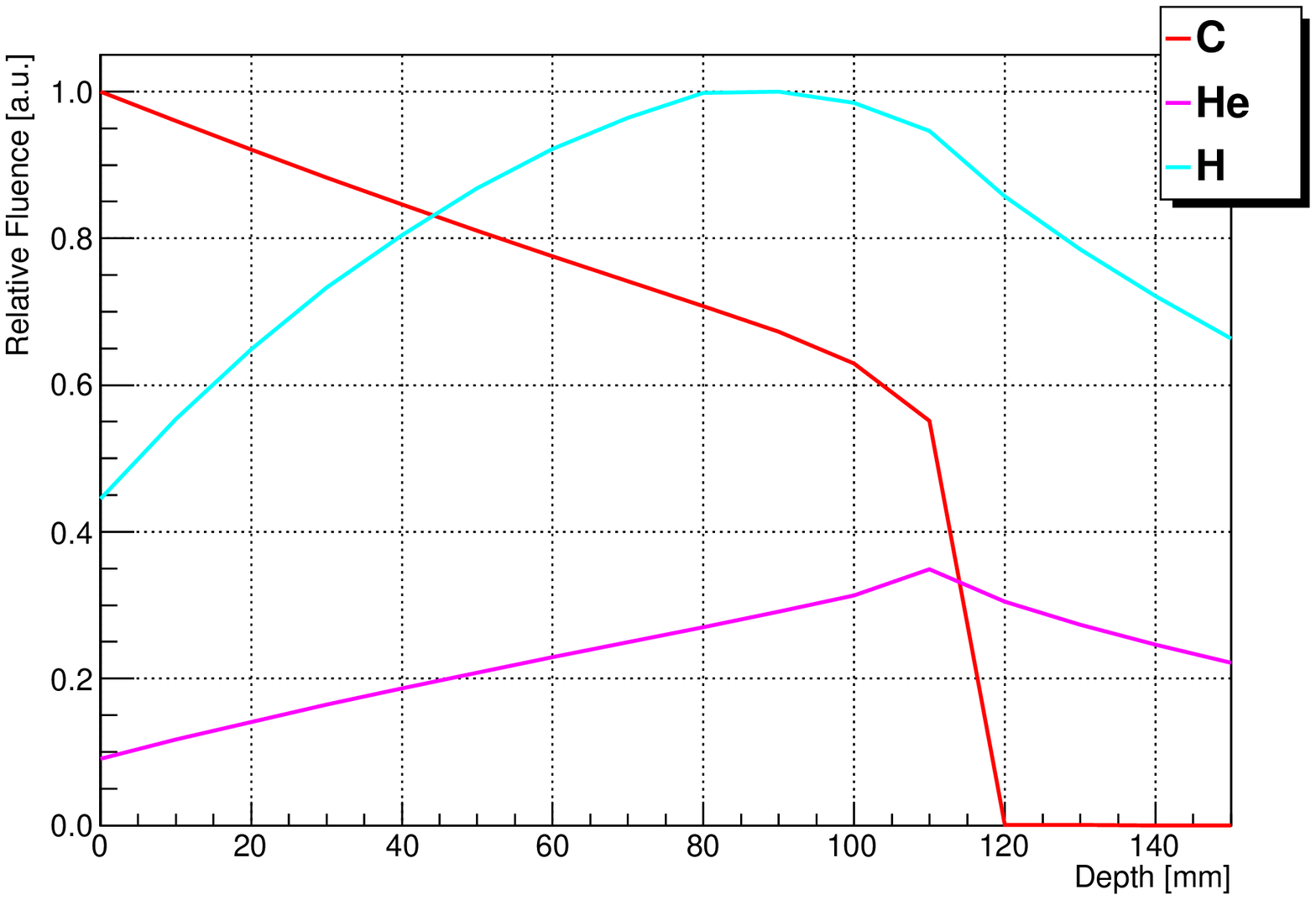}
\end{minipage}
\begin{minipage}[t]{0.5\hsize}
\includegraphics[width=7.5cm]{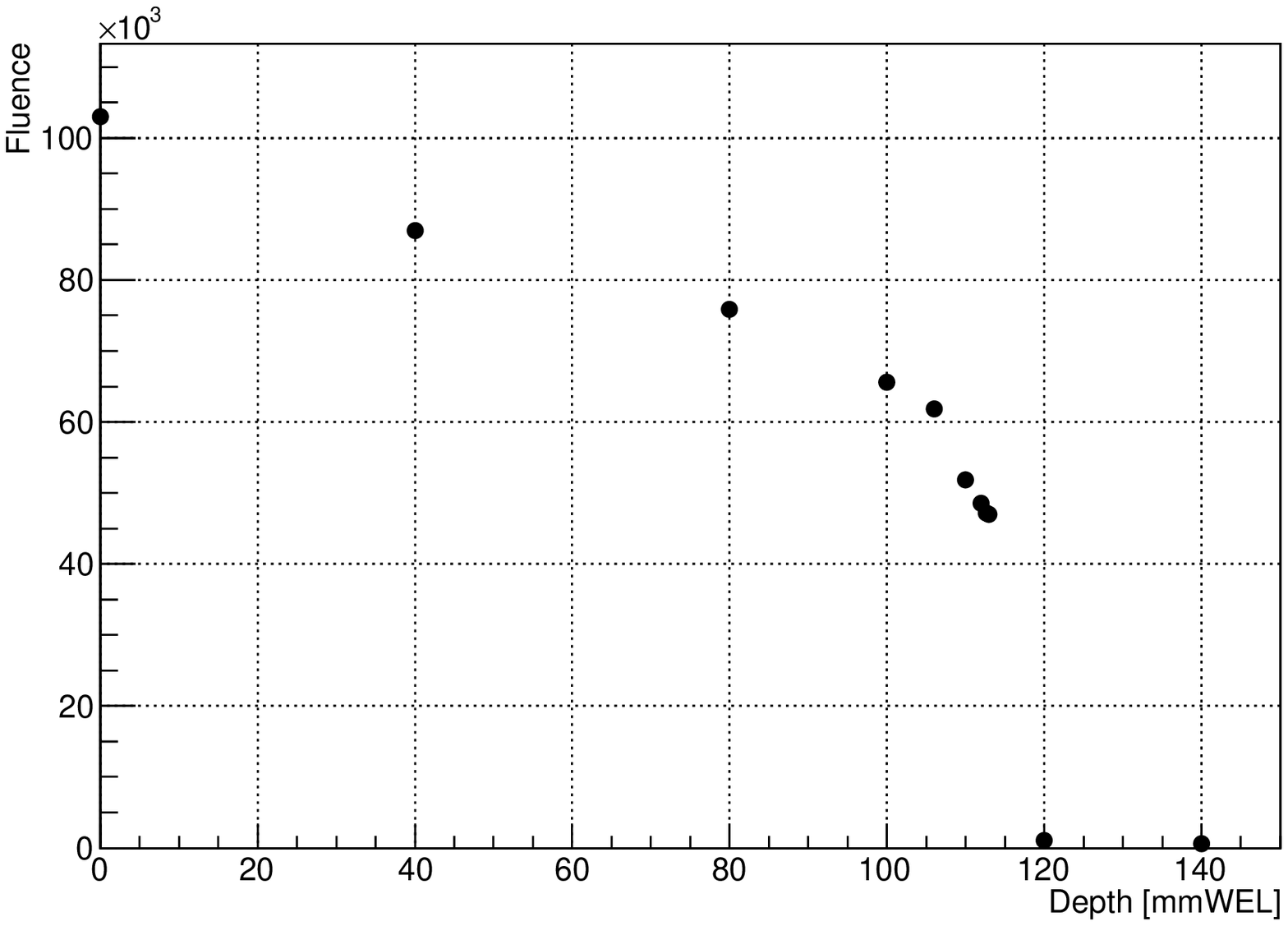}
\end{minipage}
\end{tabular}
\caption{Fluence as a function of water depth. Left shows the
simulation result for primary carbon and major secondaries, helium
and proton. Right represent experimental result for carbon candidates.
WEL means a water equivalent length.}
\label{fig:Fluence}
\end{figure}

Another approach to investigate this major peak was performed by
the estimation of the fluence. 
For simplicity, the number of events above the threshold set to 
the local minimum, e.g. 1 V for Figure \ref{fig:1DHist} (c) and
3.2 V for Figure \ref{fig:1DHist} (d), were evaluated at each
depth. Figure \ref{fig:Fluence} shows the evaluated fluence together
with the simulation result as a function of depth in the water.
The fluence of primary carbon ions decrease in the water while 
that of the secondary particles increase due to the
nuclear reaction. The fluence of primary carbon immediately drop
down to zero at the Bragg peak.
Comparing the experimental result with simulation result, the major
peak shown in {\it Sum} distribution can be identified as primary
carbon ions.

\begin{figure}[htb]
\centering
\includegraphics[width=7.5cm]{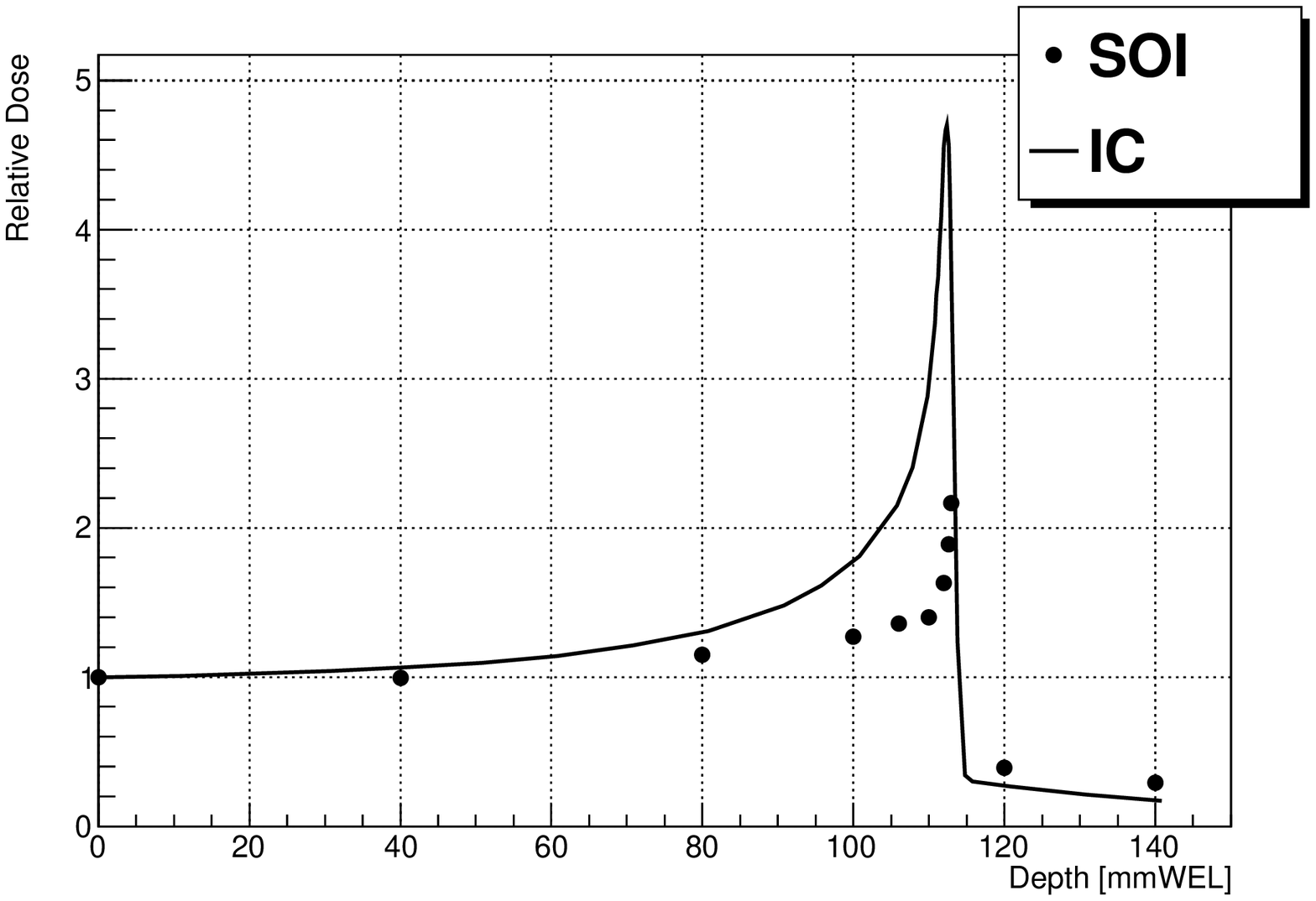}
\caption{Depth dose distribution.
The solid line and closed circle represent measurement result by 
the ionization chamber and SOPHIAS, respectively. WEL means 
a water equivalent length.}
\label{fig:PDD}
\end{figure}

Assumed that the generated electron by a radiation are corrected without
any loss, the signal of each {\it particle} should be proportional to the
amount of energy deposit. 
Thus the absorbed dose can be expressed as a sum of each signal. 
Obtained depth dose distribution normalized
at 0 mmWEL is shown in Figure \ref{fig:PDD}. 
Obviously, the signal reduction can be seen around the Bragg peak 
where the LET rapidly increase. 
In contrast, the relative dose beyond the Bragg peak is
slightly higher than the result by the ionization chamber. 
These phenomena indicate the LET dependence of SOPHIAS detector.

\begin{figure}[htb]
\centering
\includegraphics[width=7.5cm]{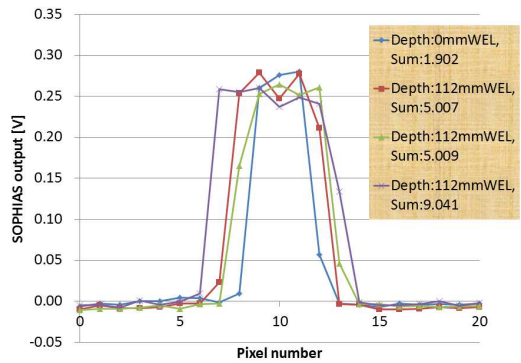}
\caption{Signal profiles for various depths and {\it Sum} values.}
\label{fig:Profile}
\end{figure}

Figure \ref{fig:Profile} shows some signal profiles for various 
depths and {\it Sum} values. 
It is found that the shape is not similar and maximum output seems
to be about 0.25 V. 
It should be noted that the this output together with the pedestal
( $<$0.2 V ) is much smaller than the dynamic range. Though it is
difficult to explain this phenomenon with this study,
the recombination effects might be one possible candidate.
More than a hundred million electron-hole pairs are populated
in the 500 $\mu$m thick Si by the carbon ion at Bragg peak region.
They could be diffused and recombined during the process of collection.
The device simulator study is therefore necessary to understand 
this phenomenon.

\clearpage
\section{Summary}
The experimental study regarding the response of SOI image sensor,
SOPHIAS, to the therapeutic carbon ion beam was carried out.
It was confirmed that the primary carbon can be identified with
a high spatial resolution. This will be a great advantage for
the precise quality assurance and R$\&$D of new irradiation system.
The gain reduction was found around the high LET region.
This LET dependence will be studied by using device simulator in
the near future.

\Acknowledgements
The author express his gratitude to Dr. Y. Arai, Dr. T. Hatsui,
Dr. T. Kudo and all of the C02 group members for giving me
an opportunity for this research and useful discussions.
This work was supported by JSPS KAKENHI Grant Number 26109501.

\end{document}

%% file: econfmacros.tex
\def\beq{\begin{equation}}
\def\eeq#1{\label{#1}\end{equation}}
\def\eeqn{\end{equation}}

\def\beqa{\begin{eqnarray}}
\def\eeqa#1{\label{#1}\end{eqnarray}}
\def\eeqan{\end{eqnarray}}

\let\bar=\overbar

\def\Dslash{\not{\hbox{\kern-4pt $D$}}}
\def\dslash{\not{\hbox{\kern-2pt $\del$}}}

\def\msb{{\bar{\ssstyle M \kern -1pt S}}}




%% file: eprint.bbl
\begin{thebibliography}{99}


\bibitem{Tsujii}
H. Tsujii, and T. Kamada, 
"A Review of Update Clinical Results of Carbon Ion Radiotherapy", 
Jpn. J. Clin. Oncol. {\bf 42}, 670-685 (2012).

\bibitem{Ohno}
T. Ohno, 
"Particle radiotherapy with carbon ion beams",
EPMAJ, {\bf 4}, 9 (2013).

\bibitem{Kanai}
T. Kanai, M. Endo, S. Minohara, {\it et al.}, 
"Biophysical characteristics of HIMAC clinical irradiation system 
for heavy-ion radiation therapy",
Int. J. Radiat. Oncol. Biol. Phys., {\bf 44}, 201-210 (1999).

\bibitem{Torikoshi}
M. Torikoshi, S. Minohara, N. Kanematsu {\it et al.}, 
"Irradiation System for HIMAC",
J. Radiat. Res., {\bf 48}, A15-A25 (2007).

\bibitem{Ohno2}
T. Ohno, T. Kanai, S. Yamada, {\it et al.}, 
"Carbon Ion Radiotherapy at the Gunma University Heavy 
Ion Medical Center: New Facility Set-up",
Cancers, {\bf 3}, 4046-4060 (2011).

\bibitem{Arai}
Y. Arai, T. Miyoshi, Y. Unno, {\it et al.}, 
"Development of SOI pixel process technology",
NIM A, {\bf 636}, S31-S36 (2011).

\bibitem{Hatsui}
T. Hatsui, M. Omodani, T.Kudo , {\it et al.}, 
"A direct-detection X-ray CMOS image sensor with 500 $\mu$m 
thick high resistivity silicon",
Proceedings of the International Image Sensor Workshop (IISW),
12-16 (2013).

\end{thebibliography}
